\documentclass[11pt, letter]{article}
\usepackage{jheppub}
\usepackage[english]{babel}
\usepackage{blindtext}\blindmathtrue
\usepackage{avant} 
\usepackage[dvipsnames]{xcolor}
\usepackage{amsmath,epsf,amssymb,mathtools,latexsym,amsthm,setspace,array,pifont,hyperref,amsfonts,dsfont,cancel,braket,parskip,slashed}
\usepackage[none]{hyphenat} 
\usepackage{graphicx}
\usepackage{float}
\allowdisplaybreaks[1]
\makeatletter
\graphicspath{{Pictures/}} 
\usepackage{todonotes}
\newcommand{\nn}{\nonumber}
\newcommand{\sd}{\mathrm{d}}
\newcommand{\pd}{\partial}

\newcommand{\cl}[1]{\mathcal{#1}}

\def\prd{\ref@{Phys.~Rev.~D}}        

\usepackage{tcolorbox}
\definecolor{airforceblue}{rgb}{0.36, 0.54, 0.66}
\definecolor{azure}{rgb}{0.0, 0.5, 1.0}
\newtcolorbox{tdbox}{colback=airforceblue!40!white,colframe=azure!90!black} 
\newcommand{\td}[1]{
	\if\notesOn1
	\begin{tdbox}
		#1
	\end{tdbox}
	\fi
}
\hypersetup{
	colorlinks = true,
	linkcolor = Mahogany,
	anchorcolor = Mahogany,
	citecolor = Mahogany,
	filecolor = Mahogany,
	urlcolor = Mahogany
}

\def\notesOn{1}

\usepackage{enumitem,amssymb}
\newlist{todolist}{itemize}{2}
\setlist[todolist]{label=$\square$}
\usepackage{pifont}
\title{Quantization Conditions and the Double Copy}
\author[1]{William~T.~Emond,}
\author[2]{Nathan Moynihan}
\author[2]{and Liangyu Wei}
\affiliation[1]{CEICO, Institute of Physics of the Czech Academy of Sciences, Na Slovance 2, 182 21 Praha 8, Czech Republic}
\affiliation[2]{Higgs Centre for Theoretical Physics, School of Physics and Astronomy, The University of Edinburgh, EH9 3FD, Scotland}
\emailAdd{william.emond@fzu.cz}
\emailAdd{nathantmoynihan@gmail.com}
\abstract{	
	We formulate Wilson loop observables as products of eikonal Wilson lines given in terms of on-shell scattering amplitudes. We derive the eikonal phases for dyons in both gauge theory and gravity, which we use to derive the Dirac-Schwinger-Zwanziger quantization condition and its relativistic gravitational (Taub-NUT) counterpart via the double copy. We also compute the Wilson loop for an anyon-anyon system, obtaining a relativistic generalisation of the Aharonov-Bohm phase for gravitational anyons. 
}

\begin{document}
\maketitle

\section{Introduction}
The eikonal approximation \cite{Abarbanel:1969ek,Wallace:1977ae,Cheng:1969eh, Levy:1969cr,Kabat:1992tb} is currently undergoing somewhat of a renaissance in the on-shell scattering amplitudes programme, with particular focus on using it to understand gravitational dynamics. This has lead to a number of important developments in both the eikonal 
\cite{Saotome:2012vy,Akhoury:2013yua,Ciafaloni:2015xsr,Bjerrum-Bohr:2016hpa,Luna:2016idw,Bjerrum-Bohr:2018xdl,Ciafaloni:2018uwe,DiVecchia:2019kta,DiVecchia:2019myk,KoemansCollado:2019ggb,Bern:2020gjj,Parra-Martinez:2020dzs,DiVecchia:2020ymx,Mogull:2020sak,AccettulliHuber:2020oou,DiVecchia:2021bdo,DiVecchia:2021ndb,Jakobsen:2021smu,Shi:2021qsb,Heissenberg:2021tzo,Haddad:2021znf,Herrmann:2021tct,Bjerrum-Bohr:2021din,Carrillo-Gonzalez:2021mqj,Jakobsen:2021lvp} and other exponential representations of the $S$-matrix \cite{Damgaard:2021ipf,Brandhuber:2021eyq}. 
The eikonal regime typically examines $2\rightarrow 2$ scattering at large centre of mass energy and large impact parameter, in which the relativistic $S$-matrix can be conveniently resummed into an exponential, whose exponent is the eikonal phase \cite{Abarbanel:1969ek,Wallace:1977ae,Cheng:1969eh, Levy:1969cr,Kabat:1992tb,Amati:1987wq,Amati:1990xe,Ciafaloni:2014esa}. Crucially, this eikonal phase contains a great deal of the classical physics (within the eikonal regime), proving extremely useful in cases where one is interested in the scattering of a probe particle off of a massive (charged) particle, mediated by a massless gauge boson or graviton. The eikonal approach requires one to assume a separation of scales between the mass $m$ of the heavy particle, the energy $\omega$ of the massless mediator and the momentum transfer $|\mathbf{q}|$ of the scattering process, i.e. one evaluates the relevant 2-to-2 scattering amplitude in the approximation $m\gg\omega\gg|\mathbf{q}|$. Although not itself an observable, the eikonal phase provides an important connection between amplitudes and observables, encoding a number of useful quantities of a physical system. For example, changes in spin,  classical impulses, scattering angles and time advance/delay. In this paper, we will elaborate on how  one can derive certain features of Wilson loops from the relevant eikonal phase.

Inspired by the progress made within the on-shell framework using eikonal physics, as well as the developments made in the study of dualities \cite{Saotome:2012vy,Caron-Huot:2013fea,Colwell:2015wna,Huang:2019cja,Alawadhi:2019urr,Banerjee:2019saj,Moynihan:2020gxj,Moynihan:2020ejh,Emond:2020lwi,Csaki:2020inw,Terning:2020dzg,Csaki:2020yei}, in this paper we shall explore a couple of interesting set-ups in both electromagnetism (EM) and gravity, via the double copy. Working at leading-order in the eikonal approximation, we shall consider configurations in EM in which the corresponding particles are dyons, which have both electric and magnetic charge. Through the established double copy of this set-up \cite{Luna:2015paa,Huang:2019cja,Emond:2020lwi,Alfonsi:2020lub,Bahjat-Abbas:2020cyb,Kim:2020cvf}, we can then extend this analysis to its gravitational analogue in a Taub-NUT background. Moreover, one can consider analogous systems in $2+1$-dimensions, in which the objects of interest are anyons and their gravitational counterparts. We shall further demonstrate how one can broaden the eikonal approach to 3D (making contact with \cite{Burger:2021wss}), and in each of the 4D and 3D scenarios, derive the associated eikonal phases, from which we will provide and illustrative example, namely the leading-order classical impulse, of how one can efficiently extract physical observables.
	
While the eikonal phase is not itself an observable, this is not true of all phases: the phase associated to a Wilson loop, whether in gauge theory or gravity, is observable via the Aharonov-Bohm effect. In a recent paper, two of the present authors connected the Aharonov-Bohm phase in 2+1 dimensions to an on-shell scattering amplitude, and in this paper we extend this to consider Wilson loops, expressing the phase directly in terms of the eikonal phase. To test these ideas, we use the formulated Wilson loops to examine quantization in gauge theory and, via the double copy, in gravity. The quantization conditions are typically derived by considering the Aharonov-Bohm phase associated to a Dirac string, demanding that the phase is undetectable since the Dirac string ought to be unobservable. Usually this involves imposing the single-valuedness of the wave function or by demanding that two potentials defined on different regions of space have a well-defined overlap. However, in the spirit of the on-shell framework, here we shall adopt  a different approach, with the simple (and reasonable) requirement that observables built out of on-shell scattering amplitudes be gauge invariant. This line of enquiry has often led to interesting insights, for example much of the physics of black holes is very simply captured by some relatively simple on-shell amplitudes \cite{Arkani-Hamed:2019ymq,Moynihan:2019bor,Emond:2019crr,Chung:2018kqs,Guevara:2018wpp,Guevara:2019fsj,Burger:2019wkq}.
	
There have been some very interesting recent papers on Wilson loops (and Wilson lines) and the double copy \cite{White:2011yy,Alfonsi:2020lub,Alawadhi:2021uie}, typically focussing on the double copy at the level of the fields. Here, we adopt a different approach, formulating the Wilson loop entirely in terms of on-shall amplitudes, which we can apply the double copy to directly. We make use of the fact that the eikonal phase itself corresponds to a Wilson line, which enables us to construct a Wilson loop by ``gluing" together these lines in a prescribed manner, or in other words, summing two eikonal phases evaluated along a closed path. We demonstrate the effectiveness of this approach in deriving the well known Dirac-Zwanziger-Schwinger quantization condition in electromagnetism and, through the double copy, we extend this to its gravitational counterpart,  enforcing the relativistic quantization of energy in terms of an associated `NUT' charge. This establishes a key result of our paper: that charge and energy quantization can be realised from a purely on-shell perspective, simply by demanding the gauge invariance of the associated Wilson loop. Finally, one can again expand the analysis to $2+1$-dimensions, within which we determine the relevant Wilson loops and in the non-relativistic, Chern-Simons limit, recover the results found for the Aharonov-Bohm phase in \cite{Burger:2021wss}.

\section{Scattering Amplitudes and Eikonal Physics}
We start by defining the eikonal phase for a four-particle amplitude $\mathcal{A}_4[p_1,p_2\rightarrow p_1',p_2']$. In momentum space, such an amplitude can be expressed purely in terms of masses $M_i$ and the Mandelstam invariants $s$ and $t$.
By now, it has become standard to express the eikonal phase in terms of the transverse Fourier transform
\begin{equation}\label{key}
	e^{\frac{i}{\hbar}\chi} -1 = i\int \hat{\sd}^Dq\,\hat{\delta}(2p_1\cdot q)\,\hat{\delta}(2p_2\cdot q)\,e^{iq\cdot b}\mathcal{A}_4[p_1,p_2\rightarrow p_1',p_2']\big\vert_{q^2\rightarrow 0} \;.
\end{equation}
The eikonal phase $\chi$ can be expanded in powers of coupling constant, encompassing various loop contributions from the amplitude, albeit with a number of constraints from unitarity. In general one has to be careful in applying the eikonal phase at higher order, where the quantum remainder can in principle contribute, however in this paper we will only be interested in leading order effects, i.e.
\begin{equation}\label{eikonaldef}
	\chi = \frac{\hbar}{4M_1M_2}\int\hat{\sd}^Dq\,\hat{\delta}(u_1\cdot q)\hat{\delta}(u_2\cdot q)\,e^{iq\cdot b}\mathcal{A}_4[p_1,p_2\rightarrow p_1',p_2']\big\vert_{q^2\rightarrow 0} \;.
\end{equation}

As claimed, one can readily extract the leading-order impulse of one of the particles, say particle 1, taken to be a probe in the presence of a much heavier particle (in this case, particle 2), from the eikonal phase $\chi$. This is achieved by taking the derivative of $\chi$ with respect to the projected impact parameter: $\Pi^\mu_{\;\;}\frac{\pd}{\pd b_\nu}$. The projector is a dimensionally-dependent quantity, and in four dimensions is given by 
\begin{equation}\label{key}
	\Pi^\mu_{\;\;\nu}=(\gamma^2 -1)^{-1}\epsilon^{\mu\rho\alpha\beta}\epsilon_{\nu\rho\gamma\delta}u_{1\alpha}u_{2\beta}\,u_1^{\gamma}u_2^{\delta}\;,
\end{equation}
whereas in three dimensions by
\begin{equation}\label{key}
	\Pi^\mu_{\;\;\nu}=(\gamma^2 -1)^{-1}\epsilon^{\mu\rho\alpha}\epsilon_{\nu\rho\gamma}u_{1\alpha}u_2^{\gamma}\;.
\end{equation}
In both cases, the projector ensures that the resulting expression stays
transverse to both the incoming velocities $u_1^\mu$ and $u_2^\mu$ meaning its components lie in the impact parameter plane. As such, we can express the leading-order impulse as
\begin{equation}\label{eq:impulse from eikonal}
	\Delta p_1^\mu  =  \Pi^\mu_{\;\;\nu}\frac{\pd \chi}{\pd b_\nu}  =  \frac{\hbar}{4M_1M_2}\Pi^\mu_{\;\;\nu}\frac{\pd}{\pd b_\nu}\int\hat{\sd}^4q\,\hat{\delta}(u_1\cdot q)\hat{\delta}(u_2\cdot q)\,e^{iq\cdot b}\mathcal{A}_4[p_1,p_2\rightarrow p_1',p_2']\big\vert_{q^2\rightarrow 0} \;.
\end{equation}
Having given a brief overview on relating scattering amplitudes to the eikonal phase, and how to extract the classical leading-order impulse from it, let us proceed by considering examples in both electromagnetic and gravitational settings. 
  
\subsection{Dyon-Dyon Eikonal Phase and the Electromagnetic Impulse}
Here, we wish to determine the eikonal phase associated to a pair of spinning dyons with masses $M_i$ and momentum $p_i^\mu=M_i u_i^\mu$ with $i = 1,2$. To do so, we first need to determine the relevant four-particle scattering amplitude. We will suppress factors of $\hbar$ for much of the calculations, restoring it when useful. 

The three-particle amplitude of an electromagnetically charged spin-$s$ particle with mass $M_1$ emitting a massless particle of helicity $\pm 1$ is given by \cite{Arkani-Hamed:2017jhn}
\begin{equation}\label{key}
	\mathcal{A}_{3}\left[1^s,1^{s\prime}, q^{\pm 1}\right]= \sqrt{2}eM_{1} x_1^{\pm}\frac{\braket{\textbf{11}'}^{2s}}{M_1^{2s}}.
\end{equation}
To construct \textit{dyon} amplitudes, we duality-rotate the above three-particle amplitudes and take the infinite spin limit, finding \cite{Emond:2020lwi}
\begin{equation}\label{key}
	\mathcal{A}_{3}\left[1,1^{\prime}, q^{\pm 2}\right]= \sqrt{2}eM_{1} x_{1}^{\pm}e^{\pm(i \theta_1+q \cdot a_1)}, \quad \mathcal{A}_{3}\left[2, 2, q^{\pm 2}\right]=\sqrt{2}e M_{2} x_{2}^{\pm} e^{\pm(i \theta_2+q \cdot a_2)} \;,
\end{equation}
where $a_1^\mu$ and $a_2^\mu$ parametrise the (classical) spins of particles 1 and 2, respectively.

The four-particle scattering amplitude is then given by
\begin{align}\label{eq:EM 4point}
	\cl{A}_4 &= \frac{2M_1M_2e^2}{q^2}\left(\frac{x_1}{x_2}e^{-i(\theta_1-\theta_2)-q(a_1+a_2)} + \frac{x_2}{x_1}e^{i(\theta_1-\theta_2)+q(a_1+a_2)}\right) \nn\\[0.5em]
	&= \frac{2M_1M_2}{q^2}\Big[(e_1-ig_1)(e_2+ig_2)\left(u_1\cdot u_2 + i\frac{\epsilon(\eta,u_1,q,u_2)}{q\cdot\eta}\right)e^{-iq\cdot a} \nn\\ &\qquad\qquad\, + (e_1+ig_1)(e_2-ig_2)\left(u_1\cdot u_2 - i\frac{\epsilon(\eta,u_1,q,u_2)}{q\cdot\eta}\right)e^{iq\cdot a}\Big] \;,
\end{align}
where we have defined $a \equiv -i(a_1+a_2)$, along with the duality-rotated couplings
\begin{equation}\label{key}
	e_i \equiv e\cos\theta_i,~~~~~ g_i \equiv e\sin\theta_i. 
\end{equation}
The dyon-dyon eikonal phase is then given by
\begin{flalign}
	\chi_{_{\text{dyon}}}  &= \frac{1}{2}\int\hat{\sd}^4q\,\hat{\delta}(u_1\cdot q)\hat{\delta}(u_2\cdot q)\frac{e^{iq\cdot b}}{q^2}\Big[(e_1-ig_1)(e_2+ig_2)\left(u_1\cdot u_2 + i\frac{\epsilon(\eta,u_1,q,u_2)}{q\cdot\eta}\right)e^{-iq\cdot a} \nn\\ &\qquad\qquad\qquad\qquad\qquad\qquad\quad\, + (e_1+ig_1)(e_2-ig_2)\left(u_1\cdot u_2 - i\frac{\epsilon(\eta,u_1,q,u_2)}{q\cdot\eta}\right)e^{iq\cdot a}\Big]\nn
\end{flalign}
\begin{equation}
	= \ \text{Re}\,(e_1-ig_1)(e_2+ig_2)\int\hat{\sd}^4q\,\hat{\delta}(u_1\cdot q)\hat{\delta}(u_2\cdot q)\frac{e^{iq\cdot ( b-a)}}{q^2}\left(u_1\cdot u_2 + i\frac{\epsilon(\eta,u_1,q,u_2)}{q\cdot\eta}\right)\;.
\end{equation}
To extract the leading-order probe particle impulse from the eikonal phase, we need simply refer back to eq.~\eqref{eq:impulse from eikonal}, from which we can determine the following result: 
\begin{flalign}\label{eq:dyon impulse}
	\Delta p_1^\mu \ =&\  \Pi^\mu_{\;\;\nu}\frac{\pd}{\pd b_\nu}\chi_{_{\text{dyon}}} \nn\\[0.5em] =& \ \text{Re}\,i(e_1-ig_1)(e_2+ig_2)\int\hat{\sd}^4q\,\hat{\delta}(u_1\cdot q)\hat{\delta}(u_2\cdot q)\,\frac{e^{iq\cdot ( b-a)}}{q^2}\left(q^\mu\,u_1\cdot u_2 - i\epsilon^\mu(u_1,q,u_2)\right)  \;,
\end{flalign}
where we have used that $q^\mu\epsilon(\eta,u_1,q,u_2)=-(q\cdot\eta)\epsilon^\mu(u_1,q,u_2) +\mathcal{O}(q^2)$. Eq.~\eqref{eq:dyon impulse} exactly corresponds to the dyon-dyon generalisation of the result found in \cite{Emond:2020lwi}, firmly establishing the applicability of the eikonal phase, and its efficiency at extracting classical observables.

\subsection{The Gravitational Eikonal Phase and Impulse via the Double Copy}\label{ssec:gravity eikonal impulse}
We can of course carry out the same procedure in a gravitational setting. From an amplitudes perspective, one can simply apply the double copy to the electromagnetic case, mapping the spinning dyon scattering amplitude to that of two interacting duality-rotated Kerr black holes (Kerr-Taub-NUT's, cf.~\cite{Emond:2020lwi}). Thus, the corresponding set-up in gravity, is to consider two duality-rotated Kerr black holes: a probe of mass $M_1$ and momentum $p_1^\mu$, interacting with a heavy black hole of mass $M_2$($\gg M_1$) and momentum $p_2^\mu$. Up to a constant of proportionality, the relevant three-point gravitational amplitude $\mathcal{M}_3$ is simply the square of $\mathcal{A}_3$, and thus upon a duality rotation, and taking the infinite spin limit, we find
\begin{equation}\label{key}
	\mathcal{M}_{3}\left[1,1^{\prime}, q^{\pm 2}\right]=\frac{\kappa}{2}M_{1}^2 x_{1}^{\pm 2}e^{\pm(i \theta_1+q \cdot a_1)}, \quad \mathcal{M}_{3}\left[2, 2, q^{\pm 2}\right]=\frac{\kappa}{2} M_{2}^{2} x_{2}^{\pm 2} e^{\pm(i \theta_2+q \cdot a_2)},
\end{equation}
where $\kappa^2 = 32\pi G$. The long-range amplitude that we are interested in is therefore given by
\begin{align}\label{eq:gravity 4-point}
	\mathcal{M}_{4} &= \left(\frac{\kappa}{2}\right)^{2} \frac{M_{1}^{2} M_{2}^{2}}{q^{2}}\left(\left(\frac{x_{1}}{x_{2}}\right)^{2} e^{i (\theta_1-\theta_2)+q \cdot (a_1 + a_2)}+\left(\frac{x_{2}}{x_{1}}\right)^{2} e^{-i(\theta_1-\theta_2)-q \cdot (a_1 + a_2)}\right) \nn\\ &= \frac{2M_1M_2}{q^2}\Big[(Q_1+i\tilde{Q}_1)(Q_2-i\tilde{Q}_2)\Big(u_1\cdot u_2+i\frac{\epsilon(\eta,u_1,q,u_2)}{q\cdot\eta}\Big)^2e^{iq\cdot a} \nn\\ &\qquad\qquad\, +(Q_1-i\tilde{Q}_1)(Q_2+i\tilde{Q}_2)\Big(u_1\cdot u_2-i\frac{\epsilon(\eta,u_1,q,u_2)}{q\cdot\eta}\Big)^2e^{-iq\cdot a}\Big] \;,
\end{align}
where in analogy to the electromagnetic case we have introduced a new set of dual couplings
\begin{equation}\label{dualCouplings}
	Q_i = \frac{\kappa M_i}{2\sqrt{2}}\cos\theta_i = \sqrt{4\pi G} m_i\;,~~~~~\tilde{Q}_i = \frac{\kappa M_i}{2\sqrt{2}}\sin\theta_i = \sqrt{4\pi G} \ell_i\;,
\end{equation}
where $m_i$ and $\ell_i$ are the mass and dual mass (i.e. the mass rescaled by $\cos\theta$ or $\sin\theta$).

As an interesting aside, note that amplitude \eqref{eq:gravity 4-point} is invariant under \textit{double} duality, that is to say, if we scatter two Kerr-Taub-NUT objects ($Q_1 = Q_2 = 0$) or two Kerr black holes ($\tilde{Q}_1 = \tilde{Q}_2 = 0$), the scattering amplitude is identical, up to coupling labels. This reflects the fact that Hodge duality acting twice is trivial. The interesting physics arises when we consider the scattering of a Kerr black hole off a Kerr-Taub-NUT, i.e. $\tilde{Q}_1 = Q_2 = 0$, or two gravitational dyons.

Returning to the problem at hand, using \eqref{eq:gravity 4-point} we determine the gravitational eikonal phase to be
\begin{flalign}
	\chi_{_{\text{KTN}}} &= \frac{1}{2}\int\hat{\sd}^4q\,\hat{\delta}(u_1\cdot q)\hat{\delta}(u_2\cdot q)\frac{e^{iq\cdot b}}{q^2}\Big[(Q_1+i\tilde{Q}_1)(Q_2-i\tilde{Q}_2)\Big(u_1\cdot u_2+i\frac{\epsilon(\eta,u_1,q,u_2)}{q\cdot\eta}\Big)^2e^{iq\cdot a} \nn\\ &\qquad\qquad +(Q_1-i\tilde{Q}_1)(Q_2+i\tilde{Q}_2)\Big(u_1\cdot u_2-i\frac{\epsilon(\eta,u_1,q,u_2)}{q\cdot\eta}\Big)^2e^{-iq\cdot a}\Big] \nn\\[0.5em] &= \text{Re}\,(Q_1+i\tilde{Q}_1)(Q_2-i\tilde{Q}_2)\int\hat{\sd}^4q\,\hat{\delta}(u_1\cdot q)\hat{\delta}(u_2\cdot q)\frac{e^{iq\cdot (b+a)}}{q^2}\Big(u_1\cdot u_2+i\frac{\epsilon(\eta,u_1,q,u_2)}{q\cdot\eta} \Big)^2 \nn\\[0.5em] &= \text{Re}\,(Q_1+i\tilde{Q}_1)(Q_2-i\tilde{Q}_2)\int\hat{\sd}^4q\,\hat{\delta}(u_1\cdot q)\hat{\delta}(u_2\cdot q)\frac{e^{iq\cdot (b+a)}}{q^2} \nn\\ &\qquad\qquad\qquad\qquad\qquad\qquad\qquad \times\Big[\cosh(2w)+2i\frac{(u_1\cdot u_2)\epsilon(\eta,u_1,q,u_2)}{q\cdot\eta}\Big] \;,
\end{flalign}
where in going from the second to the third equality, we have expanded the square, and noted that
\begin{equation}\label{eq:levi-civita expansion}
	\epsilon^2(\eta,u_1,q,u_2)=-(q\cdot\eta)((u_1\cdot u_2)^2 -1)+\mathcal{O}(u_i\cdot q)+\mathcal{O}(q^2) \;,
\end{equation} 
in which we can ignore terms proportional to $q\cdot u_1$ or $q\cdot u_2$, noting that these do not contribute to the Wilson loop due to the presence of the delta functions. Moreover, we neglect contact terms $\mathcal{O}(q^2)$ since this are set to zero in the definition of the eikonal phase. Finally, we have used that $u_i^2=-1$ ($i=1,2$), and $u_1\cdot u_2=-\gamma=-\cosh(w)$, such that $2(u_1\cdot u_2)^2-1=\cosh(2w)$, where $w$ is the rapidity.

As in the electromagnetic case, it is then a simple procedure to extract the leading-order impulse of the the probe Kerr-Taub-NUT from $\chi_{_{\text{KTN}}}$. Indeed, using eq.~\eqref{eq:impulse from eikonal}, the result is
\begin{flalign}\label{eq:ktn impulse}
	\Delta p_1^\mu \ =&\  \Pi^\mu_{\;\;\nu}\frac{\pd}{\pd b_\nu}\chi_{_{\text{KTN}}} \nn\\[0.5em] =& \ \text{Re}\,(Q_1+i\tilde{Q}_1)(Q_2-i\tilde{Q}_2)\int\hat{\sd}^4q\,\hat{\delta}(u_1\cdot q)\hat{\delta}(u_2\cdot q)\,\frac{e^{iq\cdot (b+a)}}{q^2}\nn\\ &\qquad\qquad\qquad\qquad\qquad\qquad\qquad\times\left(iq^\mu\cosh(2w)-2\cosh(w)\epsilon^\mu(u_1,q,u_2)\right)  \;,
\end{flalign}
where we have used that $q^\mu\epsilon(\eta,u_1,q,u_2)=(q\cdot\eta)\epsilon^\mu(u_1,q,u_2) +\mathcal{O}(q^2)$. This result is again in exact agreement with that found in \cite{Emond:2020lwi}. Of course, this is to be expected, but is nevertheless an elegant and efficient way to derive the classical impulse, bearing in mind that the eikonal phase encodes a number of physical observables. Indeed, we shall shortly move on to discuss how (in certain cases) Wilson loops can be derived, given knowledge of the appropriate eikonal phases. This is the subject of section~\ref{sec:Wilson loops}. Before moving on, however, we shall now briefly elaborate on how this approach can be extended to study analogous physical scenarios in 2+1 dimensions.

\subsection{Anyon-Anyon Eikonal Phase and the Electromagnetic and Gravitational Impulse}
Given the recent renewed interest in gauge theories and gravity in 2+1 dimensions, here we shall extend our analysis to the corresponding configurations in this case: anyons and their gravitational counterpart. In doing so, we are able to make contact with some of the results derived in \cite{Burger:2021wss}. The efficiency that the eikonal phase approach affords us, means that the results can be obtained in only a few lines.

The four particle amplitude in the small $m$ limit is given by
\begin{equation}
		\cl{A}_4[1,2,1',2'] = \frac{2e_1e_2M_1M_2}{q^2+m^2}\left(u_1\cdot u_2 + i\frac{m\epsilon(u_1,u_2,q)}{q^2} + \frac{m^2}{4m_1m_2}\right)\label{osamp} \;,
\end{equation}
such that the three-dimensional electromagnetic eikonal phase is then
\begin{equation}\label{EMAnyonChi}
	\chi_{_{\text{anyon}}} = \frac{e_1e_2}{2}\int\hat{\sd}^3q\,\hat{\delta}(u_1\cdot q)\hat{\delta}(u_2\cdot q)\,\frac{e^{iq\cdot b}}{q^2+m^2}\left(u_1\cdot u_2 + i\frac{m\epsilon(u_1,u_2,q)}{q^2}\right) \;.
\end{equation}
For gravitational anyons, the four particle amplitude given by \cite{Burger:2021wss}
\begin{equation}\label{key}
	\cl{M}_4[1,2,1',2'] = 2\kappa^2\frac{m_1^2 m_2^2 }{q^2+m^2}\left(\cosh 2w + 2iu_1\cdot u_2 \frac{m\epsilon(u_1, u_2, q)}{q^2} \right) - 2\kappa^2\frac{m_1^2 m_2^2\sinh^2w}{q^2} \;,
\end{equation}
giving an eikonal phase
\begin{align}\label{GRAnyonChi}
	\chi_{_{\text{grav anyon}}} &= \frac{\kappa^2 m_1m_2}{2}\int\hat{\sd}^3q\,\hat{\delta}(u_1\cdot q)\hat{\delta}(u_2\cdot q)\,\frac{e^{iq\cdot b}}{q^2+m^2}\left(\cosh 2w + 2iu_1\cdot u_2 \frac{m\epsilon(u_1, u_2, q)}{q^2}\right)\nn\\
	&\quad -  \frac{\kappa^2 m_1m_2}{2}\int\hat{\sd}^3q\,\hat{\delta}(u_1\cdot q)\hat{\delta}(u_2\cdot q)\,\frac{e^{iq\cdot b}}{q^2}\sinh^2w \;.
\end{align}

As a check, we can derive the impulses in both cases, as the eikonal phases will be used later on. The electromagnetic impulse is given by
\begin{flalign}\label{eq:anyon_impulse}
	\Delta p_1^\mu \ =&\  \Pi^\mu_{\;\;\nu}\frac{\pd}{\pd b_\nu}\chi_{_{\text{anyon}}} \nn\\[0.5em] =& \ \frac{ie_1e_2}{2}\int\hat{\sd}^3q\,\hat{\delta}(u_1\cdot q)\hat{\delta}(u_2\cdot q)\,\frac{e^{iq\cdot b}}{q^2+m^2}\left(q^\mu u_1\cdot u_2 - im\epsilon^\mu(u_1,u_2)\right) \;,
\end{flalign}
where we have used 
\begin{equation}
	q^\mu\epsilon(q,u_1,u_2) = -m^2\epsilon^\mu(u_1,u_2) + \cl{O}(m^2/m_1) \;.
\end{equation}
In the gravity case, we find that the impulse is 
\begin{flalign}\label{eq:grav_anyon impulse}
	\Delta p_1^\mu \ =&\  \Pi^\mu_{\;\;\nu}\frac{\pd}{\pd b_\nu}\chi_{_{\text{grav anyon}}} \nn\\[0.5em] =& \ \frac{\kappa^2}{2}m_1m_2\int \hat{\sd}^3q~\hat{\delta}(u_1\cdot q)\hat{\delta}(u_2\cdot q)\,e^{-iq\cdot b}\nn\\
	&~~~~~~~~~~~~~~~~~~~~~~~~~~\times\left(\frac{q^\mu\cosh2w}{q^2+m^2}-\frac{2im\epsilon^\mu(u_1, u_2)\cosh w}{q^2+m^2} - \frac{q^\mu\sinh^2w}{q^2}\right) \;.
\end{flalign}
Upon inspection, it is found that both the electromagnetic and gravitational results are again in exact agreement with those determined previously in \cite{Burger:2021wss}, affirming the validity of the eikonal phase in 2+1 dimensions.

\section{Wilson Loops and Scattering Amplitudes}\label{sec:Wilson loops}
  
The Wilson loop is an observable quantity in gauge theory, derived from the holonomy of the gauge connection around some closed contour $\cl{C}$. It essentially corresponds to a phase factor, but is observable via the Aharonov-Bohm effect and thus physical. Generically, it is defined as the trace of the path-ordered exponential of a gauge field $A_\mu$ integrated around a closed spacetime contour $\mathcal{C}$
\begin{equation}\label{eq:wilson loop def}
	W[A_\mu] = \text{Tr}\left(\mathcal{P}\exp\left(\frac{ie}{\hbar}\oint_{\mathcal{C}} A_\mu(x) \sd x^\mu \right)\right) \;,
\end{equation}
where $\mathcal{P}$ is the path-ordering operator.

In the case of electromagnetism, the gauge field is $U(1$) and thus Abelian, meaning that eq.~\eqref{eq:wilson loop def} reduces to  
\begin{equation}\label{key}
	W[A_\mu] = \exp\left(\frac{ie}{\hbar}\oint_{\mathcal{C}} A_\mu(x) \sd x^\mu \right) = \exp\left(\frac{ie}{\hbar}\oint_{\mathcal{C}} A_\mu(x(\tau)) \frac{\sd x^\mu}{\sd\tau}\sd\tau \right),
\end{equation}
where $\tau$ is an affine parameter such that contour is given by $\cl{C} = \left\{x^\mu = x^\mu(\tau), \tau \in [0,1]\right\}$.

The Wilson loop is gauge invariant, and in this analysis we wish to calculate it directly from an on-shell scattering amplitude. To do so, we will first break our contour $\cl{C}$ into two paths, $\cl{C} = \gamma_+ + \gamma_-$, where, to leading order, each path corresponds to evaluating $x^\mu$ as a straight line. This breaks the Wilson loop into a product of Wilson lines, however we need to be careful to ensure that our two chosen paths do indeed produce a spacetime loop, ensuring that any amplitudes we consider describe scattering in the same plane.

We will choose the two paths to be determined by $x^\mu_1 = u_1^\mu\tau + b^\mu$ and $x^\mu_2 = -u_1^{\mu}\tau - b^\mu$ where $u_1^\mu \equiv \frac{p_1^\mu}{M_1}$ is the proper velocity of a particle probing the potential generated by a particle with momentum $p_2^\mu$, which sources the electromagnetic (or gravitational) potential. 

\begin{figure}[H]
	\centering
	\includegraphics{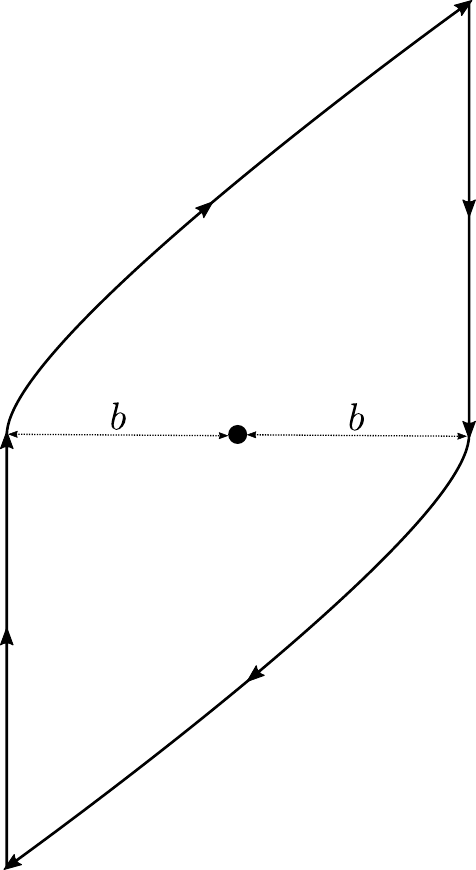}
	\caption{Closed contour as a product of Wilson Lines.}\label{fig:closedContour}
\end{figure}

At $\tau \rightarrow\infty$, the proper velocity picks out points on the celestial sphere, since
\begin{equation}\label{key}
	\frac{x^\mu_1}{\sqrt{-x^2_1}} \bigg|_{\tau \rightarrow \pm \infty} = \pm u_1^\mu = -\frac{x^\mu_2}{\sqrt{-x^2_2}}\bigg|_{\tau \rightarrow \pm \infty} \;.
\end{equation}
In Fig.~\ref{fig:closedContour} we present a diagram of how this closed contour is constructed from the above argument.

As such, the Wilson loop can be expressed as
\begin{align}\label{key}
	W[A_\mu] &= \exp\left(\frac{ie}{\hbar}\oint_{\gamma_1 + \gamma_2} A_\mu(x(\tau)) \frac{\sd x^\mu}{\sd\tau}\sd\tau \right) \nn\\[0.5em]&= \exp\left(\frac{ie}{\hbar}\int_{-\infty}^\infty \sd\tau\,\left[A_\mu(u_1\tau + b) u_1^\mu - A_\mu(u_1\tau - b) u_1^{\mu}  \right]\right).
\end{align}
There is, however a caveat here, in that this is only \textit{strictly} valid for attractive potentials, since this always ensures that the paths will have crossed by the time they reach asymptotic infinity. In principle, for a potential that vanishes as $r\rightarrow\infty$, any two paths could be connected by a translation (or rotation) at asymptotic infinity, where the gauge field is zero. However, this has the potential to introduce effects arising from large gauge transformations and care is likely needed. In this paper, we will only consider attractive potentials and so we will not comment on this further.

To recast this in terms of an on-shell scattering amplitude, we first recall a well-known quantity from relativistic electromagnetism, namely the relation between the gauge field $A^\mu$ and the electric potential $V$. Indeed, for a relativistic charged point particle subject to an external electromagnetic field, the potential is given by 
\begin{equation}\label{key}
	V(p_1,q) = e\,u_1\cdot A(q) = \frac{e}{M_1}p_1\cdot A(q) \;.
\end{equation}
With this expression in hand, we can simply employ the (relativistic) Born approximation\footnote{Such a potential can be derived using old-fashioned perturbation theory via the Lippmann-Schwinger equations. Interestingly, it is possible to derive the relativistic eikonal approximation in this manner \cite{Todorov:1970gr}.} to relate the potential $V$ to a corresponding 4-point scattering amplitude $\cl{A}_4$. At leading order, it can be shown that they are related by
\begin{equation}\label{key}
	\braket{p|V|p+q} = \frac{\cl{A}_4(p,p+q)}{4M_1M_2} \;,
\end{equation} 
where $\cl{A}_4(p,p+q)$ is the $2\rightarrow 2$ scattering amplitude. 

We can use this relation to write the Wilson loop, at leading order, as
\begin{align}\label{eq:wilson loop}
	W[A_\mu] &= \exp\left(\frac{1}{4M_1M_2\hbar}\int_{-\infty}^\infty d\tau\int \hat{\sd}^Dq\,\hat{\delta}(u_2\cdot q)\,\cl{A}_4[p_1,p_2\rightarrow p_1',p_2']e^{iq\cdot u_1\tau}\left(e^{-iq\cdot b} - e^{iq\cdot b}\right)\right)\nn\\[0.5em]
	&= \exp\left(-\frac{i}{2M_1M_2\hbar}\int \hat{\sd}^Dq\,\hat{\delta}(u_1\cdot q)\,\hat{\delta}(u_2\cdot q)\,\cl{A}_4[p_1,p_2\rightarrow p_1',p_2']\sin(q\cdot b)\right).
\end{align}
We recognise the phase as a sum of two eikonal phases defined in eq. \eqref{eikonaldef}, each evaluated at different values of the impact parameter $b$.

It is useful to note that any part of the amplitude which is an even function of $q^2$ will not contribute to the Wilson loop, since the integral will vanish. This is because only the spatial part of $q$ is non-zero (due to the delta functions) and is therefore an even function such that
\begin{equation}\label{eq:parity symmetric}
	\int \hat{\sd}^4q\,\hat{\delta}(u_1\cdot q)\,\hat{\delta}(u_2\cdot q)\,f(q^2)\sin(q\cdot b) = 0 \;.
\end{equation}

\subsection{Quantization of Electric Charge from Amplitudes}
Long ago, Dirac showed that if a single magnetic monopole with charge $g$ exists in the universe, then electric charge $e$ must be quantized in terms of $g$ via \cite{Dirac:1931kp}
\begin{equation}\label{key}
	eg = 2\pi\hbar n \;,
\end{equation}
with $n$ an integer\footnote{Note that we have restored $\hbar$ in this expression, and do so where necessary throughout this paper, such that the quantum and classical results extracted from the eikonal phases and Wilson loops are clearly distinguished.}. This was further generalised independently by both Schwinger \cite{Schwinger:1966nj} and Zwanzinger \cite{Zwanziger:1968ams, Zwanziger:1968rs} to include particles with both electric \textit{and} magnetic charge - dyons - which leads to the constraint
\begin{equation}\label{key}
	e_1 g_2 - e_2 g_1 = 2\pi \hbar n \;,
\end{equation}
known as the generalised Dirac-Zwanziger-Schwinger quantization condition. This was also extended to the gravitational case by Dowker and Roche \cite{Dowker:1967zz} (see also \cite{Misner:1963fr, 1974GReGr...5..603D}), where in the non-relativistic limit it was found that gravitational mass $m$ is quantized in terms of the NUT charge $\ell$ as
\begin{equation}\label{key}
	m\ell = \frac14\hbar n \;.
\end{equation} 
It was later shown by Bunster et al \cite{Bunster:2006rt}, that this could extended to a relativistic setting, resulting in the generalised condition
\begin{equation}\label{key}
	\gamma G (m_1\ell_2 - m_2\ell_1) = \frac14\hbar n \;.
\end{equation}
In this section we examine the set-ups in electromagnetism and gravity that lead to these constraints within the on-shell framework. To do so, we focus on the corresponding Wilson loops, which are by definition gauge invariant objects, observable via the Aharonov-Bohm effect. The aim is to recover the known quantization conditions by simply imposing the gauge invariance of the Wilson loop, without knowledge of the underlying geometry, or the associated wavefunction of the system in this set-up.

With this in mind, having established the connection between Wilson loops and scattering amplitudes, we shall proceed by computing the Wilson loop generated by a dyon-dyon amplitude. Let us do so in the spinless case, i.e. $a^\mu =0$, as including spin does not add anything of any particular relevance\footnote{For non-zero $a^\mu$ the Wilson loop has a more complicated structure, however, it turns out that the spin contribution is already gauge invariant, and therefore provides no further constraints. As it only serves to complicate the expression, distracting from the main result, we work in the spinless case in the analysis of this section. The argument is the same in the gravitational case.}. 

Starting from the dyon-dyon eikonal phase \eqref{eq:EM 4point} (with $a^\mu =0$), and massaging it into a form that exposes the different charge sectors, we can determine the corresponding Wilson loop, by plugging it into \eqref{eq:wilson loop}. As per eq.~\eqref{eq:parity symmetric} and the explanation preceding it, we can neglect any parts of the amplitude that are an even function of $q^2$, as their integrals vanish. As such, we can consider the reduced amplitude,
\begin{equation}
	\tilde{\cl{A}}_4 = 4M_1M_2(e_1g_2-e_2g_1)\frac{\epsilon(\eta,u_1,q,u_2)}{q^2(q\cdot\eta)} \;,
\end{equation}
which readily gives us the following Wilson loop:
\begin{flalign}\label{eq:em wilson loop}
	W[A_\mu] \ =& \ \exp\Big(\frac{1}{\hbar}(e_1g_2-e_2g_1)\epsilon(\eta,u_1,u_2)_\mu J^\mu \Big),
\end{flalign}
where the $J^\mu$ is given by (for further details of this result see appendix~\ref{sec:appen 1})
\begin{flalign}\label{eq:J int}
	J^\mu &= -2i\int \frac{\sd^4q}{(2\pi)^2}\delta(u_1\cdot q)\delta(u_2\cdot q)\frac{q^\mu}{q^2(q\cdot\eta)}\sin(q\cdot b) = \frac{ib_\perp^\mu}{|\epsilon(\eta_\perp,u_1,u_2,b_\perp)|}\; ,
\end{flalign}
and we have noted that for $\eta_\perp\cdot u_i = 0$ and $b_\perp\cdot u_i = 0$ ($i=1,2$), one finds that $\epsilon^2(\eta_\perp,u_1,u_2,b_\perp) = -|\beta\gamma|^2\left(b_\perp^2 \eta_\perp^2 - (b_\perp\cdot \eta_\perp)^2\right)$. 

Using this result in eq.~\eqref{eq:em wilson loop},  the Wilson loop can be expressed succinctly, as 
\begin{flalign}\label{eq:wilson loop gauge dependent}
		W[A_\mu] &= \exp\left(\frac{i}{\hbar}\,(e_1g_2-e_2g_1)\text{sgn}(\epsilon(\eta_\perp,u_1,u_2,b_\perp))\right) \;,
\end{flalign}
where $\text{sgn}(x)=\frac{x}{|x|}$ is the standard signum function. Now, in its current form, eq.~\eqref{eq:wilson loop gauge dependent} clearly depends on the gauge vector $\eta^\mu$ (or at least its projection onto the impact parameter plane). However, the Wilson loop is an observable quantity and should therefore be gauge invariant. That is, it should be independent of whatever gauge vector we choose. Notice, though, that the gauge vector only appears in the signum function, and therefore, if we choose some other vector $\tilde{\eta}^\mu$, the function can differ by at most a sign from the initial gauge. Thus, the gauge invariance of $W[A_\mu]$ can be ensured by requiring that is insensitive to the change $\eta^\mu\rightarrow -\eta^\mu$. This leads us to enforce the condition
\begin{equation}\label{eq:gauge independence}
	W[A_\mu] - W[A_\mu]\Big\vert_{\eta\rightarrow -\eta} = 2i\,\text{sgn}(\epsilon(\eta_\perp,u_1,u_2,b_\perp))\sin\left(\frac{1}{\hbar}(e_1g_2-e_2g_1)\right) \overset{!}{=}  0 \;.
\end{equation}
We see that this is readily satisfied if the following constraint on the dyonic charges holds:  
\begin{equation}\label{eq:em gauge invar condition}
	e_1g_2-e_2g_1 = n\hbar\pi \;, 
\end{equation}
which is precisely the celebrated charge quantisation condition. The important point to note here, is that this result was arrived at via a fully on-shell amplitudes approach, with no mention of any geometry (or wavefunctions), which is typically relied upon in the standard field theory calculation. We have shown here, that simply by demanding the gauge invariance of the Wilson loop, a necessary requirement of it being an observable, one can derive such results purely from the (leading-order) difference of two eikonal phases of the relevant four-point scattering amplitude.

One might worry that the scenario we have chosen is perhaps \textit{too} simple in the sense that the result might depend on our choice of path for the Wilson loop. The Dirac quantization conditions are known to be path-independent, and in fact we can arrive at the same conclusion from the scattering amplitude perspective. To alleviate this concern, we give a brief but sufficient argument. Let us now consider the most general set-up possible at leading-order, that allows for a non-trivial closed path and corresponding Wilson loop. Indeed, instead of choosing to evaluate the eikonal phase at $b^{\mu}$ and $-b^{\mu}$ we choose the paths $\gamma^+$ and $\gamma^-$ such that, at leading order, they are given by $x_1^{\mu} = u_1^{\mu}\tau + b^{\mu}$ and $x^{\prime\mu}_2 = -u_1^{\prime\mu}\tau + \ell^\mu$, where $b^\mu$ and $\ell^\mu$ are two arbitrary vectors, and $u_1^\mu$ and $u_1^{\prime\mu}$ are the proper velocities of the probe particle along each of the paths comprising the loop. In this case we of course find a much more unwieldy expression for the resulting Wilson loop. However, demanding that this expression is gauge invariant, we find that the following condition must be met
\begin{align}\label{key}
	0 &= W[A_\mu] - W[A_\mu]\Big\vert_{\eta\rightarrow -\eta}\nn\\
	 &= \Big\lvert\frac{\ell_{\perp}}{b_{\perp}}\Big\rvert^\alpha\,e^{iX}\left(-1 + e^{\frac{i}{\hbar}(e_1g_2-e_2g_1)\left(\text{sgn}(\epsilon(\eta_\perp,u_1,u_2,b_\perp)) - \text{sgn}(\epsilon(\eta_\perp,u'_1,u_2,\ell_\perp))\right)}\right),
\end{align}
where $X$ is some complicated but ultimately irrelevant phase, and $\alpha$ is some equally irrelevant exponent. This results in an identical quantization condition provided the two vectors $b^\mu$ and $\ell^\mu$ are seperated by the worldline of particle two at closest approach. In this case, the two signum functions above always add together to give $\pm 2$, resulting in the same quantization conditions as before, i.e. the one given in eq. \eqref{eq:em gauge invar condition} \footnote{The difference $W[A_\mu] - W[A_\mu]\big\vert_{\eta\rightarrow -\eta}$ further reduces identically to that given in the middle equality of \eqref{eq:em gauge invar condition} in the case where $\ell^\mu = -b^\mu$ and $u_1^{\prime\mu} = u_1^\mu$.}. This is equally true in the gravitational case considered in the next section, and so from now on we will stick to considering the simple setup give in Fig. \ref{fig:closedContour}.
\subsection{Quantization of Momentum from Amplitudes}
In this section we will turn to quantization conditions in gravity. To do so, we consider the generalisation of a Wilson loop from gauge theory to the case of perturbative gravity. The gravitational Wilson loop then corresponds to the phase experienced by a scalar test particle in a gravitational field. This can be formulated in terms of the proper length of a closed curve $\mathcal{C}$ traversed by a probe particle of mass $m$, and to leading-order in the gravitational coupling $\kappa$, is given by \cite{Green:2008kj,Brandhuber:2008tf,Alfonsi:2020lub}
\begin{align}\label{eq:gravity wilson loop}
	W[h_{\mu\nu}] &= \exp\left(-\frac{i\kappa m}{2\hbar}\oint_{\mathcal{C}} h_{\mu\nu}(x(\tau)) \frac{\sd x^\mu}{\sd\tau}\frac{\sd x^\nu}{\sd\tau}\sd\tau \right) \;.
\end{align}
To see this, we note that in the full covariant theory, the Wilson loop must take the form (in the proper time parametrisation)
\begin{equation}\label{eq:full gravity wilson loop}
	\Phi[h_{\mu\nu}] = \exp\left(\frac{i m}{\hbar}\oint_{\mathcal{C}} \left(-g_{\mu\nu}\frac{\sd x^\mu}{\sd\tau}\frac{\sd x^\nu}{\sd\tau}\right)^{1/2}\sd\tau \right) \;.
\end{equation}
Expanding the metric as $g_{\mu\nu}=\eta_{\mu\nu} + \kappa h_{\mu\nu}$ and expand the square-root, we find that to leading-order in $\kappa$, the integrand becomes
\begin{equation}\label{eq:proper length expansion}
	\left(-g_{\mu\nu}\frac{\sd x^\mu}{\sd\tau}\frac{\sd x^\nu}{\sd\tau}\right)^{1/2} = 1 - \frac{\kappa}{2}h_{\mu\nu}\frac{\sd x^\mu}{\sd\tau}\frac{\sd x^\nu}{\sd\tau} + \mathcal{O}(\kappa^2)\;.
\end{equation}
Inserting this expansion back into eq.~\eqref{eq:full gravity wilson loop} and absorbing the constant term (independent of $h_{\mu\nu}$) into an overall normalisation, we immediately recover eq.~\eqref{eq:gravity wilson loop}.

From a scattering amplitudes perspective, however, the setup is identical to the electromagnetic case, where now we simply replace the photon exchange amplitude with a graviton exchange amplitude, i.e. 
\begin{flalign}\label{key}
	W[h_{\mu\nu}] &= \exp\left(\frac{1}{4M_1M_2\hbar}\int \hat{\sd}^4q\,\hat{\delta}(u_1\cdot q)\,\hat{\delta}(u_2\cdot q)\,\cl{M}_4[p_1,p_2\rightarrow p_1',p_2']\left(e^{-iq\cdot b} - e^{iq\cdot b}\right)\right) \nn\\[0.8em] &= \exp\left(-\frac{i}{2M_1M_2\hbar}\int \hat{\sd}^4q\,\hat{\delta}(u_1\cdot q)\,\hat{\delta}(u_2\cdot q)\,\cl{M}_4[p_1,p_2\rightarrow p_1',p_2']\sin(q\cdot b)\right) \;.
\end{flalign}
To arrive at the above expression, we have used the fact that the perturbative expansion of the relativistic Lagrangian for a point particle, subject to an external gravitational field, is proportional to eq.~\eqref{eq:proper length expansion} (with the proportionality constant being the mass of the probe particle). Furthermore, as in the electromagnetic case, to leading order in the Born approximation we can relate the relativistic momentum space potential (experienced by the test particle) with the four-point graviton exchange amplitude (in the probe particle limit). Thus, for a probe particle of mass $m_1$ and momentum $p_1$, we have
\begin{equation}\label{key}
	V(p_1,q) = \frac{\kappa}{2M_1} h_{\mu\nu}(q)p_1^\mu p_1^\nu = \frac{\cl{M}_4[p_1,p_2\rightarrow p_1',p_2']}{4M_1M_2} \;.
\end{equation}
We will consider the scattering of two duality-rotated Kerr black holes (Kerr-Taub-NUT's), the established double copy of the spinning dyon \cite{Emond:2020lwi}. The relevant long-range four-point scattering amplitude in this case is the same as that used in section~\ref{ssec:gravity eikonal impulse}, i.e., eq.~\eqref{eq:gravity 4-point}. As in the electromagnetic case, adding the classical spin does not change the story at all in terms of the quantization condition, and we therefore set $a^\mu =0$ in the following analysis. This effectively reduces the problem to the case of two Taub-NUT's (previously explored in the non-relativistic setting \cite{Dowker:1967zz,Misner:1963fr, 1974GReGr...5..603D}).

To simplify things, as in the discussion on eikonal phases, we can make use of eq.~\eqref{eq:levi-civita expansion}, as well as making the different charge sectors apparent. This enables us to instead consider a simplified amplitude of the form
\begin{align}\label{key}
	\cl{M}_4 &= \frac{2M_1M_2}{q^2}\Big[(Q_1Q_2+\tilde{Q}_1\tilde{Q}_2)\cosh(2w)+ 2(Q_1\tilde{Q}_2-Q_2\tilde{Q}_1)\frac{(u_1\cdot u_2)\epsilon(\eta,u_1,q,u_2)}{q\cdot\eta}\Big] \;.
\end{align}
It is interesting to note that the Schwarzschild part of the amplitude is an even function of $q^2$, and therefore, from eq.~\eqref{eq:parity symmetric}, it cannot contribute to the Wilson loop. Consequently, we can completely neglect this term, and focus our attention on the parity odd term, proportional to the Levi-Civita tensor. The Wilson loop under consideration then becomes
\begin{align}\label{key}
	W[h_{\mu\nu}] &= \exp\Big(\frac{1}{\hbar}(Q_1\tilde{Q}_2-Q_2\tilde{Q}_1)(u_1\cdot u_2)\epsilon_\mu(\eta,u_1,u_2)\,J^\mu\Big) \;,
\end{align}
where $J^\mu$ is given by eq.~\eqref{eq:J int}. Accordingly, the gravitational Wilson loop is given by
\begin{align}\label{key}
	W[h_{\mu\nu}] &= \exp\Big(\frac{i}{\hbar}\,(Q_1\tilde{Q}_2-Q_2\tilde{Q}_1)(u_1\cdot u_2)\text{sgn}\big(\epsilon(\eta_\perp,u_1,u_2,b_\perp)\big)\Big) \nn\\[0.5em] &= \exp\Big(i\frac{4\pi G}{\hbar}\,\gamma(m_1\ell_2-m_2\ell_1)\text{sgn}\big(\epsilon(\eta_\perp,u_1,u_2,b_\perp)\big)\Big) \;,
\end{align}
where we have made the identification $Q_i\tilde{Q}_j = 4\pi Gm_i\ell_j$, where $m_i$ is the mass and $\ell_j$ the NUT parameter (see eq. \eqref{dualCouplings}). In order for this to be gauge invariant, as it should be as an observable, it should be independent of the gauge vector $\eta^\mu$. Following the same approach as the electromagnetic case and demanding that
\begin{equation}\label{key}
	W[h_{\mu\nu}] - W[h_{\mu\nu}]\bigg|_{\eta \rightarrow -\eta} = 2i\text{sgn}\big(\epsilon(\eta_\perp,u_1,u_2,b_\perp)\big)\sin\left(\frac{4\pi G}{\hbar}\,\gamma(m_1\ell_2-m_2\ell_1)\right) = 0
\end{equation}
 gives the generalized relativistic quantization condition
\begin{equation}\label{key}
	\gamma G (m_1\ell_2 - m_2\ell_1) = \frac14\hbar n \;.
\end{equation}
Recognising  $p_i^\mu = m_i u_i^\mu$ and defining a dual `magnetic' momentum $k_i^\mu = \ell_i u_i^\mu$, we can go further and note that we find a quantization condition for relativistic momentum
\begin{equation}\label{key}
	G(p_1\cdot k_2 - p_2\cdot k_1) = \frac14\hbar n \;.
\end{equation}
We see then that it is \textit{energy} that is really quantized, and not the mass as in the non-relativistic case first considered by Dowker and Roche \cite{Dowker:1967zz}. In fact, what we have found is precisely the relativistic quantization condition discovered by Bunster et al in Ref \cite{Bunster:2006rt}, where they found the same quantization condition by considering the Riemann tensor and its dual together with electric and magnetic stress energy tensors. Interestingly, in Ref. \cite{Argurio:2008zt} the authors discovered that the magnetic momentum $k_\mu$ must be present in the superalgebra if supersymmetric systems are to respect gravitational duality.

The appearance of \textit{momentum} in the quantization condition is not entirely surprising from the perspective of the double copy, which relates colour degrees of freedom (in this case U(1)) with kinematic degrees of freedom: products of non-abelian colour charges are replaced with products of kinematic objects. Here we see exactly this enacted, with products of charge being replaced by dot products of momentum.

\subsection{Electromagnetic \& Gravitational Phases for Anyons}
We now return to the study of anyons in electromagnetism and gravity, this time to compute the associated Wilson loops. In the electromagnetic case, we plug in the eikonal phase \eqref{EMAnyonChi} into the Wilson loop formula for $D = 3$ to find
\begin{align}\label{key}
	W[A_\mu] &= \exp\left(\frac{ie_1e_2m}{\hbar}\int\hat{\sd}^3q\,\hat{\delta}(u_1\cdot q)\hat{\delta}(u_1\cdot q)\,\sin(q\cdot b)\frac{\epsilon(u_1,u_2,q)}{q^2(q^2+m^2)}\right) \nn\\[0.5em]
	&= \exp\left(i\frac{e_1e_2}{2\hbar m}\left(1-e^{-m|b_\perp|}\right)\text{sgn}(\epsilon(u_1,u_2,b_\perp))\right) \;,
\end{align}
where we have retained only the parity violating term since this is the only piece that has a non-zero contribution to the Wilson loop (with the result \eqref{eq:parity symmetric} and the corresponding discussion carrying directly over to the 3D case). We have also made use of the following result:
\begin{align}\label{key}
	\int \hat{\sd}^3q ~\hat{\delta}(q\cdot u_1)\hat{\delta}(q\cdot u_2)\frac{q^\mu}{q^2(q^2+m^2)}\sin(q\cdot b) &= -\frac{i}{2|\beta\gamma|m^2}(1-e^{m|b_\perp|})\frac{b_\perp^\mu}{|b_\perp|} \nn\\[0.5em]
	&= \frac{(1-e^{m|b_\perp|})b_\perp^\mu}{2m^2\sqrt{\epsilon(u_1,u_2,b_\perp)^2}}  \;.
\end{align}
In the gravitational case, we plug in eq. \eqref{GRAnyonChi} to find
\begin{align}\label{key}
	W[h_{\mu\nu}] &= \exp\left(\frac{i\kappa^2 m_1m_2 m(u_1\cdot u_2)}{\hbar}\int\hat{\sd}^3q\,\hat{\delta}(u_1\cdot q)\hat{\delta}(u_1\cdot q)\,\sin(q\cdot b)\frac{\epsilon(u_1, u_2, q)}{q^2(q^2+m^2)}\right) \nn\\[0.5em]
	&= \exp\left(\frac{\kappa^2 m_1 m_2(u_1\cdot u_2)}{2\hbar m}\left(1-e^{-m|b_\perp|}\right)\text{sgn}(\epsilon(u_1,u_2,b_\perp))\right) \;.
\end{align}
Unlike in the four-dimensional case, there is no gauge dependence here - the sign of the phase is entirely determined by physical quantities, and so this phase should be entirely observable. Furthermore, we find that in the non-relativistic large $m$ Chern-Simons limit (keeping $e/m$ fixed), we recover the well known Aharonov-Bohm (AB) phase for both the anyon \cite{Jackiw:1989nq, Deser:1990ve} and gravitational anyon \cite{Deser:1989ri,Ortiz:1991gx,Deser:1990ve}. Arguably, this approach in obtaining the AB phase is advantageous over that adopted in our previous study of anyons~\cite{Burger:2021wss}, since the corresponding Wilson loop is efficiently calculated and contains more information. 

\section{Discussion}
In this paper, we have employed the eikonal approximation to explore a number of interesting properties of dyons in electromagnetism, and their gravitational counterparts ``Kerr-Taub-NUTs" (duality-rotated Kerr black holes) via the established double copy relating the two. In the context of on-shell scattering amplitudes, it is clear that the eikonal approach offers an efficient method for computing a number of pertinent observables, all derivable from the associated eikonal phase, itself readily calculated from a four-point tree-level amplitude.

Arguably the most interesting outcome of this analysis, is that one can use the eikonal phase to construct Wilson loops (under certain assumptions) in both electromagetism and gravity. This technique can then be used to great effect to derive the well-known Dirac-Schwinger-Zwanzinger charge quantization conditions, and via the double copy, the relativistic quantization of mass in terms of NUT charge, matching a result previously discovered by Bunster et al in Ref. \cite{Bunster:2006rt} (see also \cite{Argurio:2008nb,Argurio:2008zt}), arguably in a much more efficient and less painful manner. A particularly pleasing aspect of this is that these conditions can be derived directly from a tree-level on-shell scattering amplitude, provided we demand that gauge invariance is obeyed for observable quantities (as any good theorist should).

The usefulness of eikonal physics in on-shell scattering amplitudes framework is not limited to the realm of 4D. An interesting example in this case is the $2+1$ analogue of dyons, so-called anyons and their gravitational cousins (related via the double copy). It is a straightforward exercise to determine the associated eikonal phase for such configurations, enabling an efficient extraction of, e.g., the classical leading-order impulse, as well as a relativistic generalisation of the Aharonov-Bohm phase, afforded by the Wilson loop corresponding to this set-up. This nicely makes contact with results previously found in \cite{Burger:2021wss}.

It is clear that the application of eikonal physics and Wilson loops in the context of on-shell scattering amplitudes is a fruitful and enlightening endeavour (for other interesting examples see, e.g. \cite{Mogull:2020sak,KoemansCollado:2019ggb,AccettulliHuber:2020oou,Carrillo-Gonzalez:2021mqj,Jakobsen:2021lvp,Heissenberg:2021tzo,DiVecchia:2021bdo,DiVecchia:2020ymx,Parra-Martinez:2020dzs}), warranting continued research in the area. For example, it would be interesting to examine the double copy properties of boosted Taub-NUT solutions and their relation to gravitational shockwaves, which were recently investigated in \cite{Cristofoli:2020hnk}. This would be of particular relevance, particularly in light of upgrades to LIGO and future experiments such as LISA, which will be sensitive to the expected associated gravitational memory effects of such systems.

Another intriguing extension, would be to consider the non-Abelian case in which colour is added in the gauge theory sector. Indeed, there has already been some interesting research into classical Yang-Mills observables from amplitudes, for example, the colour impulse (as an analogue to momentum impulse) \cite{delaCruz:2020bbn}. It would be instructive to explore these sorts of structures further in the context of eikonal phases and Wilson loops. This would of course require more care in constructing the closed contour in the Wilson loop, since in this case path ordering would matter. Finally, one might wonder whether or not the Wilson loop could be defined via the analytic continuation of a Wilson line, much in the same way that scattering amplitudes have been used recently to describe bound orbits \cite{Kalin:2019rwq,Kalin:2019inp,Gonzo:2021drq}.

\subsection*{Acknowledgements}
NM would like to thank Donal O'Connell for many useful discussions. NM is supported by STFC grant ST/P0000630/1. WTE is supported by the Czech Science Foundation GA\v{C}R, project 20-16531Y.
\appendix

\section{Fourier Transforms}\label{sec:appen 1}
In the analysis of this paper, we have made use of a specialised Fourier transform, in which we project the Fourier transformed function onto the impact parameter plane. Here we shall give a detailed derivation of the most general inverse Fourier transform, of which we utilise in section~\ref{sec:Wilson loops} on Wilson loops (albeit specialised to $a^\mu=0$). 

Let us begin by defining the inverse eikonal Fourier transform as
\begin{equation}\label{key}
	\cl{F}_\pm[f(q)] \equiv \int\hat{\sd}^4q\,\hat{\delta}(u_1\cdot q)\hat{\delta}(u_2\cdot q)\,e^{\pm iq\cdot b}\,f(q).
\end{equation}
The function of interest, of which we shall determine the inverse eikonal Fourier transform of, is 
\begin{equation}
	f^\mu(q,\pm a,\eta) = \frac{q^\mu}{q^2(q\cdot\eta)}\,e^{\pm iq\cdot a} \;.
\end{equation}
It helps to choose a frame here, and so we pick
\begin{equation}\label{key}
	u_1^\mu = (\gamma,\gamma\beta,0,0),~~~~~u_2 = (1,0,0,0).
\end{equation}
For simplicity, we also define $\tilde{b}_\pm^\mu = b^\mu\pm a^\mu$. such that $\tilde{b}_\pm^\mu\vert_{a^\mu= 0} = b^\mu$. Then, using a Schwinger parametrization, along with the useful result
\begin{equation}
	\int~\hat{\sd}^2q_\perp~\frac{q_\perp^\mu}{q_\perp^2}e^{\pm iq_\perp\cdot b_\perp} = \pm \frac{ib^\mu_\perp}{2\pi|b_\perp|^2} \;,
\end{equation}
we can determine the eikonal Fourier transform, as follows:
\begin{align}\label{eq:eikonal FT}
	\cl{F}_\pm\left[f^\mu(q,\pm a,\eta)\right] &= \int\hat{\sd}^4q\,\hat{\delta}(u_1\cdot q)\hat{\delta}(u_2\cdot q)~e^{\pm iq\cdot \tilde{b}_\pm^\mu}\,\frac{q^\mu }{q^2(q\cdot\eta)} \nn\\[0.5em]
	&= -i\lim_{\epsilon\rightarrow 0}\int_0^\infty \sd\lambda\,e^{-\epsilon\lambda}\int\hat{\sd}^4q\,\hat{\delta}(u_1\cdot q)\hat{\delta}(u_2\cdot q)~\frac{q^\mu}{q^2}e^{\pm iq\cdot \tilde{b}_\pm^\mu}e^{i\lambda q\cdot\eta}\nn\\[0.5em]
	&= -i\frac{1}{|\beta\gamma|}\lim_{\epsilon\rightarrow 0}\int_0^\infty \sd\lambda\,e^{-\epsilon\lambda}\int~\hat{\sd}^2q_\perp~\frac{q_\perp^\mu}{q_\perp^2}e^{\pm iq_\perp\cdot (\tilde{b}^\mu_{_\pm,\perp}\pm\lambda\eta_\perp)}\nn\\[0.5em]
	&= \mp \frac{1}{2\pi|\beta\gamma|}\lim_{\epsilon\rightarrow 0}\int_0^\infty \sd\lambda\,e^{-\epsilon\lambda}\,\frac{\tilde{b}^\mu_{\pm,\perp}\pm \lambda\eta^\mu_\perp}{|\tilde{b}_{\pm,\perp} \pm \lambda\eta_\perp|^2} \nn\\[0.7em]
	&= \mp \alpha_{\pm}(\tilde{b}_{\pm,\perp},\eta_\perp)\,\tilde{b}^\mu_{\pm,\perp} + \beta_{\pm}(\tilde{b}_{\pm,\perp},\eta_\perp)\,\eta^\mu_\perp \;,
\end{align}
where we have defined the coefficients of $\tilde{b}_{\pm,\perp}$ and $\eta^\mu_\perp$, as 
\begin{flalign}\label{eq:beta def}
	\alpha_{\pm}(\tilde{b}_{\pm,\perp},\eta_\perp) &= \frac{1}{2\pi|\beta\gamma|}\lim_{\epsilon\rightarrow 0}\int_0^\infty \sd\lambda\,\frac{e^{-\epsilon\lambda}}{|\tilde{b}_{\pm,\perp} \pm \lambda\eta_\perp|^2} = \frac{\Big(\frac{\pi}{2} \mp 2\arctan\big(z(\tilde{b}_{\pm,\perp},\eta_\perp)\big)\Big)}{2\pi|\epsilon(\eta_\perp,u_1,\tilde{b}_{\pm,\perp},u_2)|} \;,\\[0.7em]
	\beta_{\pm}(\tilde{b}_{\pm,\perp},\eta_\perp) &= -\frac{1}{2\pi|\beta\gamma|}\lim_{\epsilon\rightarrow 0}\int_0^\infty \sd\lambda\,e^{-\epsilon\lambda}\,\frac{\lambda}{|\tilde{b}_{\pm,\perp} \pm \lambda\eta_\perp|^2} \;,
\end{flalign}
in which we have introduced the compact notation \smash{$z(\tilde{b}_{\pm,\perp},\eta_\perp)=\frac{(\tilde{b}_{\pm,\perp}\cdot\eta_\perp)}{\sqrt{\tilde{b}_{\pm,\perp}^2\eta_\perp^2-(\tilde{b}_{\pm,\perp}\cdot\eta_\perp)^2}}$}, and noted that \smash{$|\epsilon(\eta_\perp,u_1,\tilde{b}_{\pm,\perp},u_2)|=|\beta\gamma|(\tilde{b}_{\pm,\perp}^2\eta_\perp^2-(\tilde{b}_{\pm,\perp}\cdot\eta_\perp)^2)^{1/2}$}. 

We have deliberately left  $\beta_{\pm}(\tilde{b}_{\pm,\perp},\eta_\perp)$ unevaluated, as it technically contains a divergent part, and therefore would need to be further regulated. However, in this paper, such quantities are ultimately contracted with a Levi-Civita tensor of the form $\epsilon_\mu(\eta_\perp,u_1,u_2)$, and so the $\beta(\tilde{b}_{\pm,\perp},\eta_\perp)\eta^\mu_\perp$ term is completely irrelevant, as it automatically vanishes. Moreover, any observables should be independent of the gauge vector $\eta^\mu_\perp$, and as a result of these facts, we therefore do not concern ourselves further with the exact details of computing $\beta_{\pm}(\tilde{b}_\perp,\eta_\perp)$.

Returning to eq.~\eqref{eq:eikonal FT}, we see that in the case where $a^\mu= 0$, the result reduces to 
\begin{equation}
	\cl{F}_\pm\left[\frac{q^\mu}{q^2(q\cdot\eta)}\right]  = \mp \frac{\Big(\frac{\pi}{2} \mp 2\arctan\big(z(\eta_\perp,b_\perp)\big)\Big)}{2\pi|\epsilon(\eta_\perp,u_1,b_\perp,u_2)|}\,b_\perp^\mu \;.
\end{equation}
Using this result, one can readily derive the expression in \eqref{eq:J int} in the main body:
\begin{equation}
	J^\mu = -2i\left(\cl{F}_+\left[\frac{q^\mu}{q^2(q\cdot\eta)}\right] - \cl{F}_-\left[\frac{q^\mu}{q^2(q\cdot\eta)}\right]\right) = \frac{ib_\perp^\mu}{|\epsilon(\eta_\perp,u_1,b_\perp,u_2)|} \;.
\end{equation}
\bibliographystyle{JHEP}
\bibliography{refs} 
\end{document}